\documentclass[aps,prb,twocolumn, showpacs,superscriptaddress,preprintnumbers,amsmath,amssymb]{revtex4-2}

\usepackage[usenames,dvipsnames]{xcolor}

\usepackage[normalem]{ulem}
 \newcommand{\stkout}[1]{\ifmmode\text{\sout{\ensuremath{#1}}}\else\sout{#1}\fi}

\usepackage{float}
\usepackage{graphicx}
\usepackage{dcolumn}
\usepackage{bm}
\usepackage{multirow}
\usepackage{amsmath,amssymb}

\begin{document}

\title{Absence of Induced Ferromagnetism in Epitaxial Uranium Dioxide Thin Films}

\author{W. Thomas}
\affiliation{H. H. Wills Physics Laboratory, University of Bristol, Bristol, BS8 1TL, United Kingdom}

\author{F. Wilhelm}
\affiliation{European Syncrotron Radiation Facility, Grenoble C`{e}dex 9, France}

\author{S. Langridge}
\affiliation{ISIS, Rutherford Appleton Laboratory, Oxon OX11 0QX, England, United Kingdom}

\author{L. Harding}
\affiliation{H. H. Wills Physics Laboratory, University of Bristol, Bristol, BS8 1TL, United Kingdom}

\author{C. Bell}
\affiliation{H. H. Wills Physics Laboratory, University of Bristol, Bristol, BS8 1TL, United Kingdom}

\author{R. Springell}
\affiliation{H. H. Wills Physics Laboratory, University of Bristol, Bristol, BS8 1TL, United Kingdom}

\author{S. Friedemann}
\affiliation{H. H. Wills Physics Laboratory, University of Bristol, Bristol, BS8 1TL, United Kingdom}

\author{R. Caciuffo}
\affiliation{European Commission, Joint Research Centre (JRC), Postfach 2340, DE-76125 Karlsruhe, Germany}
\affiliation{Istituto Nazionale di Fisica Nucleare, Via Dodecaneso 33, IT-16146 Genova, Italy}

\author{G. H. Lander}
\affiliation{H. H.  Wills Physics Laboratory, University of Bristol, Bristol, BS8 1TL, United Kingdom}
\email{lander@ill.fr}
	\date{\today}

\begin{abstract}
Recently, Sharma \textit{et al.} [Adv. Sci. \textbf{9}, 2203473 (2022)] claimed that thin films
($\sim$ 20 nm) of UO$_2$ deposited on perovskite substrates exhibit strongly enhanced paramagnetism (called “induced ferromagnetism” by the authors). Moments of up to 3 $\mu_{\mathrm{B}}$/U atom were claimed in magnetic fields of 6 {\it T}. We have reproduced such films and, after characterisation, have examined them with X-ray circular magnetic dichroism (XMCD) at the uranium {\it M} edges, a technique that is element specific. We do not confirm the published results. We find a small increase, as compared to the bulk, in the magnetic susceptibility of UO$_2$ in such films, but the magnetisation versus field curves, measured by XMCD, are linear with field and there is no indication of any ferromagnetism. The absence of any anomaly around 30 K (the antiferromagnetic ordering temperature of bulk UO$_2$) in the XMCD signal suggests the films do not order magnetically.
\end{abstract}

\maketitle
 
\section{Introduction}
In 2022, a paper was published in Advanced Science \cite{sharma_induced_2022} claiming that for thin (20 nm) epitaxial films of UO$_2$ deposited on perovskite-type films, a large “ferromagnetic-like” signal was observed. The magnitude of the magnetic signal depended on the substrate, and corresponded to a moment of between 1.5 $\mu_{\mathrm{B}}$ to 3.5 $\mu_{\mathrm{B}}$ per uranium atom. If correct, this represents an important advance in understanding the thickness and strain-dependence of the strong antiferromagnetic interactions present in bulk UO$_2$, and would potentially open the way to possible device applications involving thin UO$_2$ films in spintronics and possible heterostructure systems involving such dioxide films \cite{dennett22}. Furthermore, the results of Ref. \cite{sharma_induced_2022}, and specifically the large induced ferromagnetic moments, cannot be understood within our present theory of the dioxide \cite{santini_multipolar_2009,zhou22}.

Using the expertise available at Bristol University \cite{springell_review_2023}, we have manufactured identical thin films, characterised them with X-rays and SQUID measurements, and then measured them with X-ray magnetic circular dichroism (XMCD) at the European Synchrotron Radiation Facility (ESRF) in Grenoble, France. This technique is element specific so focusses only on the behavior of the uranium atoms in the thin films. The results do not confirm those reported in \cite{sharma_induced_2022}. We find magnetic susceptibilities of UO$_2$ close to, but slightly higher, than those for the bulk material with linear $M/H$ curves, and a total induced moment with 17 {\it T} applied at 5 K of $\sim$ 0.3 $\mu_{\mathrm{B}}$. Further, XMCD measurements are able to determine the individual spin and orbital moments, and their ratio confirms closely to that known for bulk UO$_2$ with a $5f^2$ configuration.
We shall first describe our experiments and results, especially the XMCD, which the authors of Ref. \cite{sharma_induced_2022} did not use, and then return to a discussion of the properties of UO$_2$.

\section{Experimental procedures and results}
\subsection{Substrates and deposition}
The substrates used in the experiments were all obtained from the MTI Corporation. The three substrates were LAO (LaAlO$_3$), LSAT {(La,Sr)(Al,Ta)O$_3$}, and STO (SrTiO$_3$). These are the same as Ref. \cite{sharma_induced_2022}, with the exception of YAO (YAlO$_3$). Substrate thicknesses were 0.5 mm, single side (001) orientation polished to optical grade. Lattice parameters of the substrates were identical to tabulated values to four significant figures. However, the rocking curves (crystal mosaic) were different, both LSAT and STO were $<$ 0.06 degrees, but the LAO substrate had a rocking curve of 0.31 degrees.\\
Thin film deposition was performed using DC magnetron sputtering in a reactive gas atmosphere. This was undertaken within the dedicated actinide deposition chamber at the University of Bristol \cite{springell_review_2023}. 
In an earlier study \cite{bao_antiferromagnetism_2013}, in which we grew UO$_2$/LAO, also with a similar magnetron reactive-gas sputtering, the deposition temperature was 650 $^{\circ}$C. At such temperatures UO$_2$/LAO samples run the risk of a small distortion on cooling through the ferroelastic transition present at 560 $^{\circ}$C in LAO \cite{bao_antiferromagnetism_2013}. Our deposition temperature was lowered to 450 $^{\circ}$C to avoid any such a transition in LAO. The UO$_2$ deposition was carried out at a deposition pressure of $7.3\times10^{-3}$ mbar, with a partial oxygen pressure of $2\times10^{-5}$ mbar. The deposition rate was 0.1 nm/s for the UO$_2$. All samples were annealed to expel any excess oxygen post-growth and maintain the desired stoichiometry. This was performed prior to the deposition of a capping layer. A Nb cap of $\sim$ 10 nm was deposited on all samples on top of the UO$_2$ films. Nb was chosen due to the thin (1-2 nm) Nb$_2$O$_5$ passivisation layer that forms on its surface. These layers were confirmed by fitting the X-ray reflectivity data. In Ref. \cite{sharma_induced_2022} the films were made with pulsed laser deposition (PLD) at a temperature of 580 $^{\circ}$C.\\

\subsection{X-ray characterisation}
The first characterisation was with X-rays to measure the reflectivity and determine the film thickness. X-ray diffraction was then used to determine the position of specular and off-specular reflections allowing the lattice parameters of the UO$_2$ film to be determined. This gives a so-called {\it c} parameter along the growth direction, and in-plane, an {\it a} parameter. Since the lattice parameters of the substrate are different from those of UO$_2$, strain will be introduced into the UO$_2$ lattice, and, as we can see, the UO$_2$ films have tetragonal symmetry.All of the parameters derived from these X-ray measurements are given in Table \ref{tab:table1}, and compared to the values from Ref. \cite{sharma_induced_2022}.\\

\begin{table}
\caption{\label{tab:table1} Results from X-ray measurements of UO$_2$ films. The room temperature lattice parameter of bulk UO$_2$ is 5.471 \AA. Since the match UO$_2$/perovskite involves a rotation of 45$^{\circ}$, the number given for substrate lattice parameter is $\sqrt{2}$ $a$, where $a$ is the correct lattice parameter of the perovskite substrate. All entries except first line refer to the UO$_2$ thin films. Error bars for the strains are ±0.03\% and for lattice parameters ±0.001 \AA. The volume of the UO$_2$ unit cell = 163.76 \AA$^3$.}
\begin{ruledtabular}
\begin{tabular}{cccc}
\textrm{Substrate}&
\textrm{LAO}&
\textrm{LSAT}&
\textrm{STO}\\
Lattice parameter (\AA) & 5.358 & 5.468 & 5.521\\
\colrule
UO$_2$ film\\
Thickness (nm) Ref. \cite{sharma_induced_2022} & 19 & 22 & 21\\
Thickness (nm) Bristol& 18.3&20.3&20.6\\ \hline
Strain $\Vert$ $c$ (\%) Ref. \cite{sharma_induced_2022} & -0.15&-0.06&-0.29\\
Strain $\Vert$ $c$  (\%) Bristol &-0.33&-0.53&-0.29\\
Strain $\perp$ $c$ (\%) Ref. \cite{sharma_induced_2022} &-0.04&+0.15&+0.49\\
Strain $\perp$ $c$ Bristol (\%)&-0.60&+0.38&-0.05\\ \hline
Volume (\AA$^3$) Ref.  \cite{sharma_induced_2022} &163.39&164.14&164.89\\
Volume Bristol (\AA$^3$)&161.26&164.17&163.10\\
Vol. diff (\%) Ref. \cite{sharma_induced_2022} &-0.22&+0.24&+0.70\\
Vol. diff (\%) Bristol &-1.52&+0.26&-0.40\\

\end{tabular}
\end{ruledtabular}
\end{table}

There are clearly some differences between our samples and those of Ref. [1] for the case of strains $\perp$ {\it c} for the LAO and STO substrates. This shows already that reproducibility may not be assured. The LAO substrate, as mentioned above, had a relatively poor mosaic, so that may explain the differences for this substrate. Notably, in Ref. \cite{bao_antiferromagnetism_2013}, where LAO was also used as a substrate for a 20 nm UO$_2$ film, the values of the strains were +0.85\% $\Vert$ {\it c}, and –1.03\% $\perp$ {\it c}, giving a total volume difference of –1.21\%. In the earlier study, Ref. \cite{bao_antiferromagnetism_2013}, {\it c} $>$ {\it a}, which is found for the LAO in the present Bristol sample, but the differences were not as significant as in Ref. \cite{sharma_induced_2022}. We emphasise that whereas the actual values are not totally reproducible from one substrate to another, these differences are relativity small, and almost all strains are $<$ 1\%. However, in all cases a tetragonal symmetry is produced by depositing the UO$_2$ on the perovskite substrates. Some differences can be ascribed to the different method used for deposition of the UO$_2$ films between Ref. \cite{sharma_induced_2022} and the present work, but they are relatively small.\\
\subsection{Bulk magnetic measurements}
\begin{figure}
\begin{center}
\includegraphics[width=1.0\columnwidth]{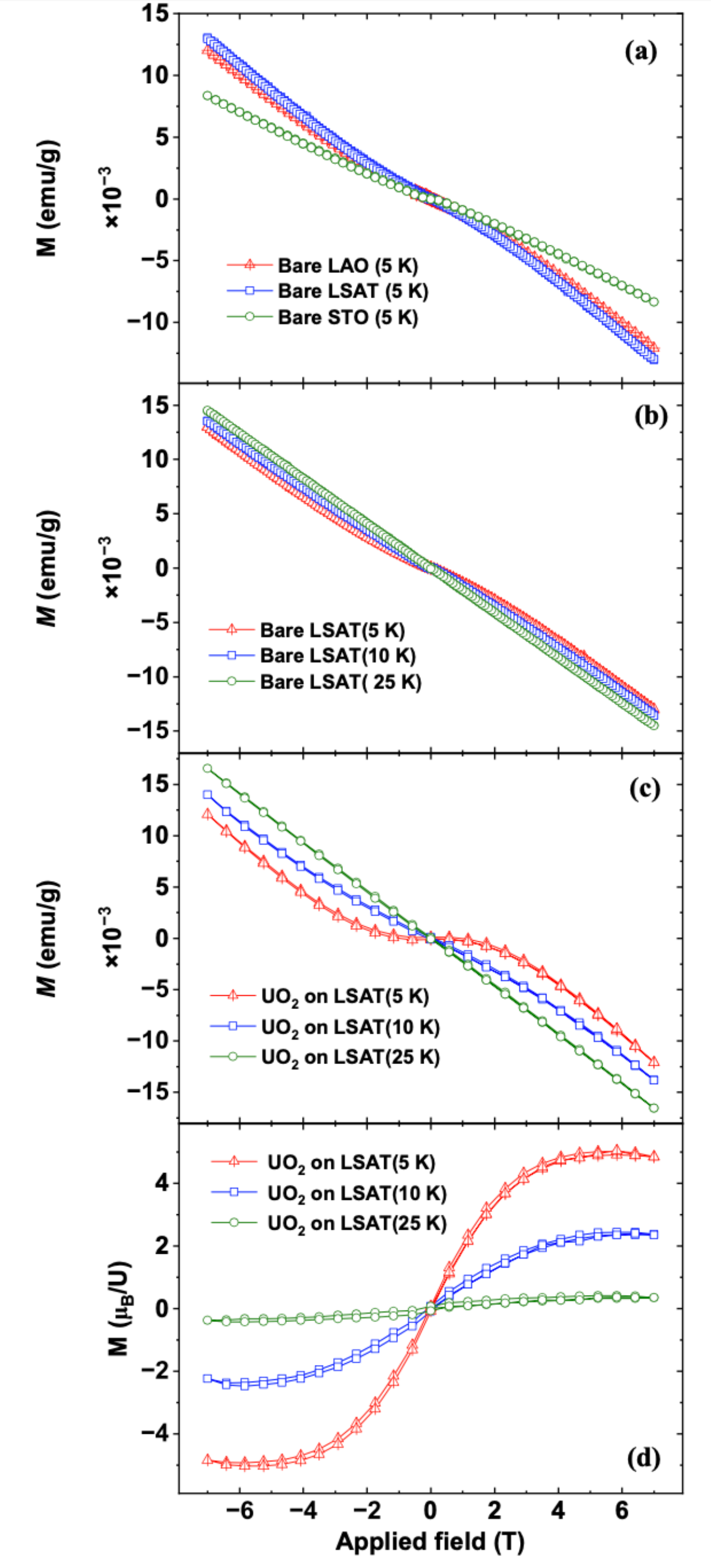}
\caption{Magnetisation results (all with $H$ in-plane) from the bare substrates (a) and (b) showing the strong diamagnetism of the substrates, as well as their temperature dependence, which is strongest at the lowest temperatures. (c) shows the results as a function of temperature for the UO$_2$/LSAT sample and (d) shows the results translated into Bohr Magnetons for the UO$_2$/LSAT sample, assuming that the bare substrates can be subtracted from the signal obtained from the substrate + film in the SQUID measurement. As we shall see, this is an incorrect assumption.}
\label{fig:1}
\end{center}
\end{figure}

We start by examining the bare substrates, i.e. before any deposition of UO$_2$ films. It is known, of course, that these substrates have a strong diamagnetic signal and that any paramagnetic signal from the UO$_2$ films will be superimposed on this strong sloping background. The results as a function of applied field in-plane are shown in Figure$.$\ref{fig:1}.\\

We note that the curves in Fig$.$ \ref{fig:1}(d) are very similar to those presented in Fig. 5(a) of Ref. \cite{sharma_induced_2022}.\\

In connection with the substrates, the work by Khalid et al. \cite{khalid_ubiquity_2010}, and by Ney et al. \cite{ney_limitations_2008} is most relevant. These authors show the precautions that are needed to make susceptibility measurements of thin films on substrates where the signals from the thin layer of interest (in this case UO$_2$) are relatively small. In particular, the work reported in Ref. \cite{khalid_ubiquity_2010} uses precisely the substrates used in the present work, LAO, LSAT, and STO, and shows in a series of figures how the substrate signal resembles a ferromagnetic response, sometimes with small coercivity, but sometimes with an appreciable value of this parameter. The authors \cite{khalid_ubiquity_2010} conclude that the simple argument of Fe impurities or other magnetic impurities (because the effects persist to relatively high temperature) is incorrect. They do not propose a final argument for why these effects are present, but argue that there is evidence that the magnetic effects may be at the surface of the substrates, rather than distributed evenly throughout the volume. The curves S3 and S4 shown in the supplementary material of Ref. \cite{sharma_induced_2022} are similar to those shown in Fig. 4, 5, 7, and 9 of Ref. \cite{khalid_ubiquity_2010}. The UO$_2$ film thicknesses in these films of Ref. \cite{sharma_induced_2022} are $\sim$ 20 nm, and the substrates have a thickness (in our case) of 0.5 mm. The ratio between these thicknesses is 25,000. If a small effect is present in the substrates and is ascribed to the films, a large and erroneous amplification is obtained.\\

\subsection{XMCD measurements}
\begin{figure}
\begin{center}
\includegraphics[width=1.0\columnwidth]{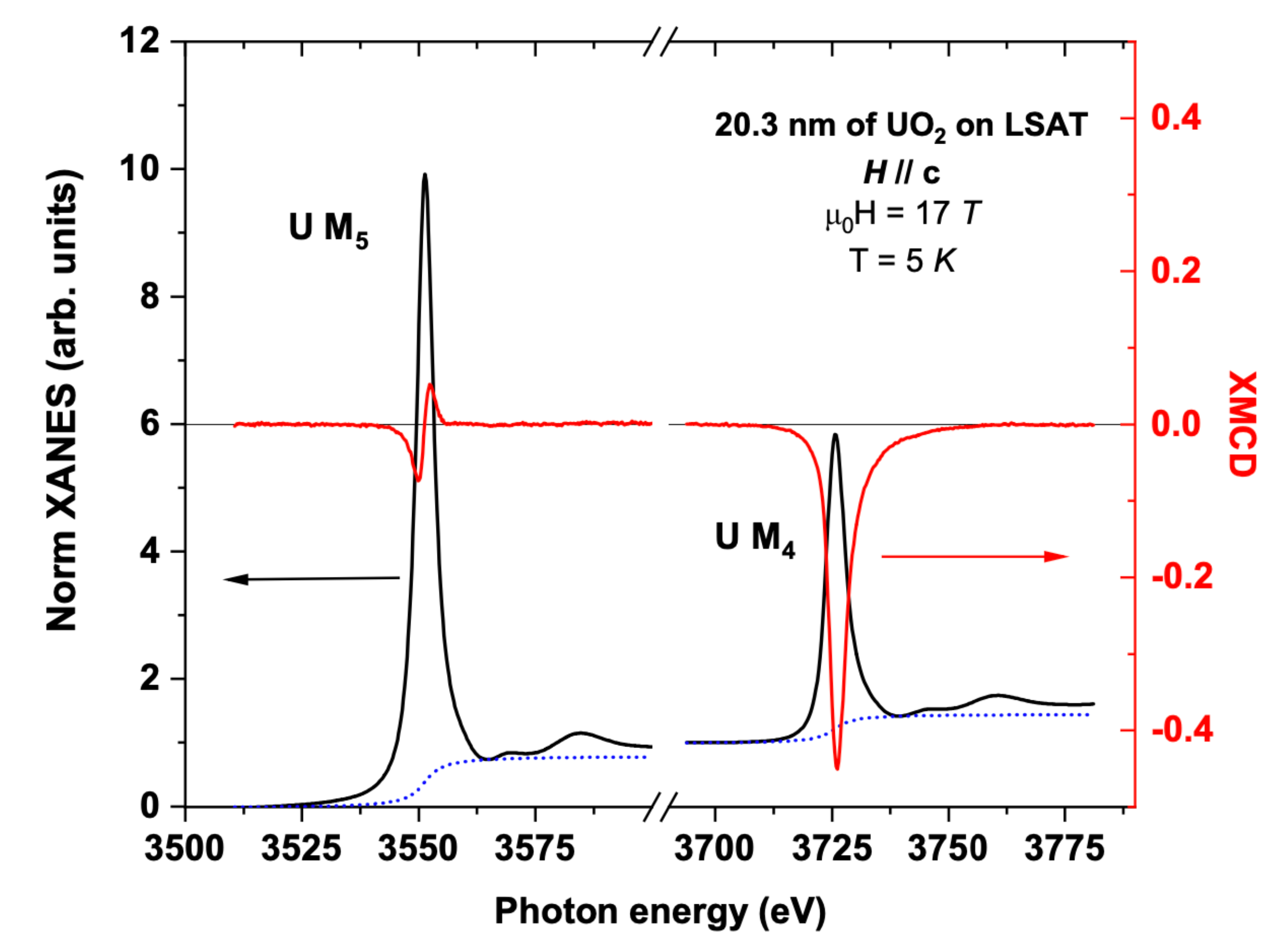}
\caption{Absorption and XMCD signals from the U $M_{4,5}$ edges of the UO$_2$/LSAT sample. The relative numbers are shown in Table \ref{tab:table2} The magnitude of the signals are similar to those found in similar work on UCoGe \cite{taupin_microscopic_2015}, although in that system the uranium is closer to $5f^3$ than UO$_2$, which is definitely $5f^2$.}
\label{fig:2}
\end{center}
\end{figure}

XMCD is a measurement that is performed at a synchrotron source, and is element specific, since the measurements are performed at an elemental absorption edge. In our case we have chosen the uranium $M_{4,5}$ edges with energies 3.73 and 3.55 keV for the spin-orbit split transition between the core {\it 3d} electrons and the unoccupied {\it 5f} states. The measurements reported below were obtained at the ID12 beamline \cite{wilhelm_magnetism_2018} at the European Synchrotron Radiation Facility (ESRF) in Grenoble, France.\\
The XMCD technique is used for most elements \cite{wilhelm_magnetism_2018}, and specifically has been useful for actinides with work on thin multilayers of U/Fe \cite{wilhelm_x-ray_2007}, and actinides as far as curium in the periodic table \cite{caciuffo_synchrotron_2023,caciuffo21}. By using the sum rules, the individual orbital and spin moments \cite{wilhelm_magnetism_2018}, \cite{caciuffo_synchrotron_2023} on the atom can be determined.
The absorption and XMCD signals are shown in Fig. \ref{fig:2}.

For H $\Vert$ {\it c}  the beam (and field) are parallel to the growth direction $[001]$, i.e. at 90$^{\circ}$ to the film. For H $\perp  c$ the field is at 10$^{\circ}$ to the plane of film and the $[100]$ normal. Self-absorption corrections are necessary for the latter measurement (as the path length can be as long as 2 mm), but are negligible for the H $\Vert$ {\it c} direction, as the average path length is only 20 nm (1/e attenuation is $\sim$ 200 nm at the $M_4$ edge and $\sim$ 120 nm at the $M_5$ edge). In all cases measured for XMCD the M vs H curves are linear, where M is the moment deduced from the sum rules, so we may plot either susceptibility or induced moment. We show in the Appendix the field-dependence of the XMCD signal for two of the samples.\\

XMCD is also an extremely sensitive technique. The mass of our 20 nm UO$_2$ films on a substrate of $5 \times 10$ mm$^2$ is $\sim$ 10 $\mu$g, and in the normal incidence configuration the beam (240 $\mu$m diameter) illuminates $\sim$ $10^{-3}$ of the sample. The technique is therefore sensitive to $\sim$ 10 ng of UO$_2$ and the dichroic signal can be readily observed in a one second per point scan across the resonant energy. Certainly, experiments can be done on thinner samples.\\

\begin{table*} 
\caption{\label{tab:table2} Values of the parameters relevant to UO$_2$ from XMCD experiments. The field applied is 17 {\it T} and the temperature is 5 K. BR is the branching ratio, which for $5f^2$ configuration should be 0.68. The error bars for UO$_2$/LAO are at least 10\% and no branching ratio could be determined due to strong scattering by the sample. The error bars for UO$_2$/LAT and UO$_2$/STO are in the 1 - 2\% range.
The final column is simply the susceptibility ($\chi$), given in $m\mu_{\mathrm{B}}/T$ and is the first column divided by the field of 17 {\it T}. In bulk UO$_2$ this number is given by Arrott \& Goldman \cite{arrott_magnetic_1957} as 1/$\chi$ = $6.5 \times 10^{-4}$ g/emu which translates to a value of $\chi$ = 7.46 $m\mu_{\mathrm{B}}/T$. The value given by Jamie {\it et al.} \cite{jaime_piezomagnetism_2017} in high-field work on bulk UO$_2$ single crystals is 7.12 $m\mu_{\mathrm{B}}/T$. }
\begin{ruledtabular}
\begin{tabular}{ccccccc}
 &Total moment&Orbital moment&Spin moment
&Orbit:Spin ratio&Branching ratio&Susceptibility\\
&($m\mu_{\mathrm{B}}$)&($m\mu_{\mathrm{B}}$)&($m\mu_{\mathrm{B}}$)& & &($m\mu_{\mathrm{B}}$) \\ \hline
UO$_2$/LAO\\
$H \Vert c$ &240&360&-120&-3.26&N/A&14.1\\
$H \perp c$ &301&439&-138&-3.18&N/A&17.7\\ \hline
UO$_2$/LSAT\\
$H \Vert c$ &297&426&-129&-3.30&0.686&17.6\\
$H \perp c$ &298&427&-129&-3.31&0.687&17.5\\ \hline
UO$_2$/STO&\\
$H \Vert c$ &301&434&-133&-3.26&0.685&17.7\\
$H \perp c$ &301&439&-138&-3.18&0.686&17.7\\ 
\end{tabular}
\end{ruledtabular}
\end{table*}

The measured values obtained from the XMCD measurements are summarised in Table \ref{tab:table2}.
Whereas, the values in the Tables are completely consistent for UO$_2$/LSAT and UO$_2$/STO samples, this is not the case for UO$_2$/LAO. This sample, where both the UO$_2$ and substrate had poor mosaics (see earlier discussion) showed multiple scattering effects that made the XMCD analysis difficult, and the values have error bars between 10 – 15\%. In contrast, the UO$_2$/LSAT and UO$_2$/STO samples, with much narrower crystal mosaic in both the substrate and UO$_2$ film gave clean signals allowing an accurate determination of the XMCD parameters (1 – 2\%). The results also show that the response of the UO$_2$ films is independent of the field orientation, as opposed to the larger H $\perp$ c values (out-of-plane response) claimed in Ref. \cite{sharma_induced_2022}.\\
Column 5 of Table \ref{tab:table2} shows the ratio between the orbital and spin moments, $\mu$(L)/$\mu$(S), which is related to the electronic structure \cite{caciuffo_synchrotron_2023}, see Table II]. We know that UO$_2$ has a $5f^2$ configuration, so that this value should be – 3.0 for Hund’s rule Russell-Saunders coupling, and for intermediate coupling should be – 3.36. Normally for uranium systems the LS Hund’s rule coupling works well, but for higher actinides, intermediate coupling is better. We see here that the difference in values is only by 10\%, and the experimental values are completely consistent with a $5f^2$ configuration, as expected from comparison with bulk UO$_2$. Another measure of the electronic configuration is the branching ratio (BR) \cite{wilhelm_magnetism_2018}, \cite{caciuffo_synchrotron_2023}, which is close to that expected \cite{wilhelm_magnetism_2018} for $5f^2$.

\subsection{Temperature dependence of the XMCD signal}
\begin{figure}
\begin{center}
\includegraphics[width=1.0\columnwidth]{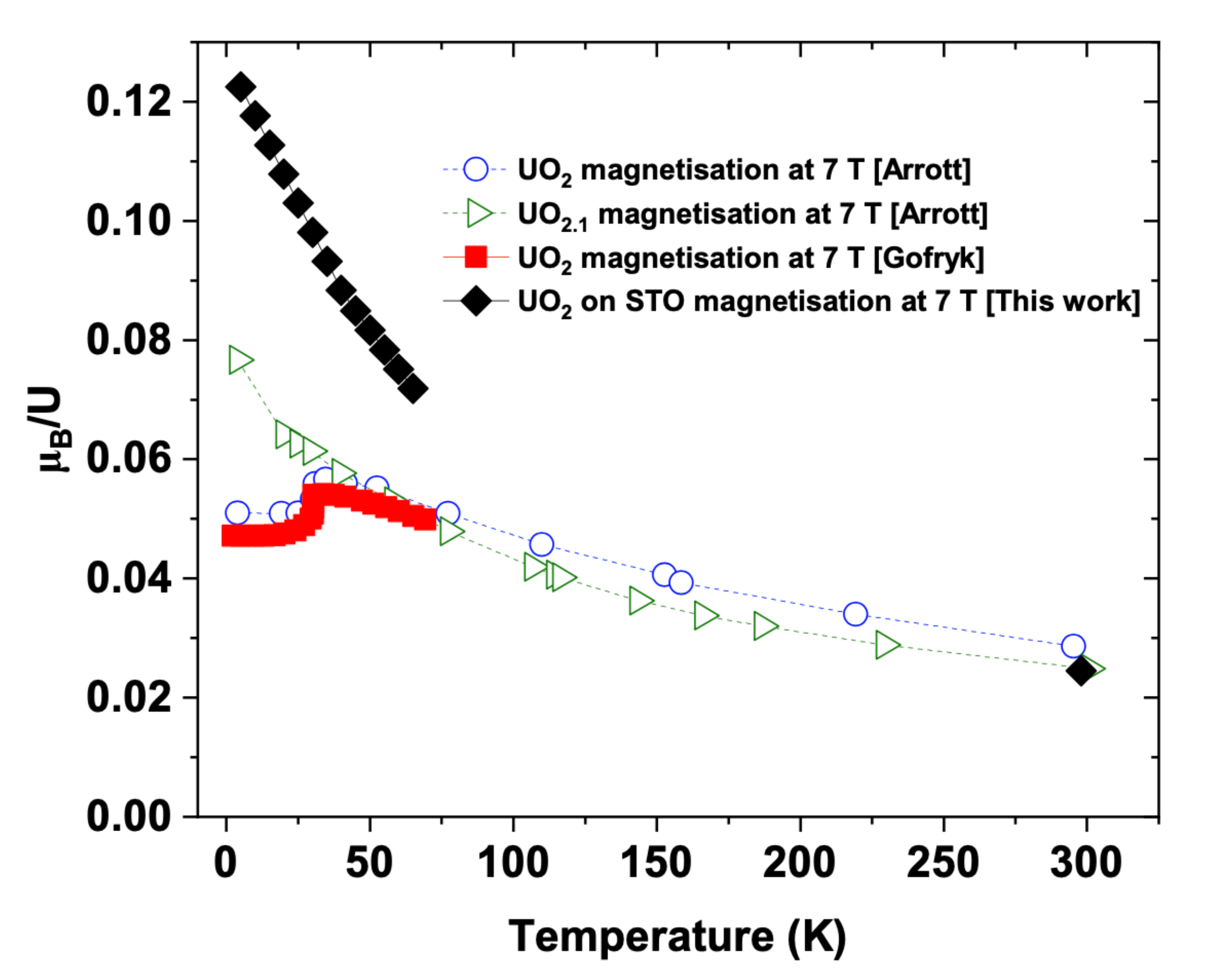}
\caption{ Moment induced on U atom as a function of temperature at H = 7 T of UO$_2$ samples. Solid symbols are from stoichiometric UO$_2$ single crystals. Black diamonds from this work. Red squares from Ref. \cite{gofryk_magnetization_2014} Open symbols from polycrystalline materials. Open blue from stoichiometric UO$_{2.0}$ and open green triangles from UO$_{2.1}$, both from Arrott \& Goldman \cite{arrott_magnetic_1957}.}
\label{fig:3}
\end{center}
\end{figure}

In the final part of our XMCD experiment we examined the signal from the UO$_2$/STO sample at the U $M_4$ edge as a function of temperature from 5 K to 60 K, and then with one point at room temperature. The applied magnetic field was fixed at 17 {\it T}. Since the integrated $M_5$ signal is small (see Fig$.$ \ref{fig:2}), the value of the XMCD signal at the $M_4$ edge is a good representation of the magnetization.\\
We plot in Figure \ref{fig:3} this value of the induced moment of a U atom scaled to a magnetic field of 7 {\it T}, as a function of temperature, together with values from the literature. All assuming a linear dependence between M and H, as shown in the Appendix.

The induced moment of our UO$_2$/STO sample is $\sim$ 2.5 times that of pure UO$_2$ at 5 K, but has the same value at 300 K. 

\section{Discussion}
Uranium dioxide has been the subject of much research since at least the 1950s. We know from work by Arrott \& Goldman \cite{arrott_magnetic_1957} that the low-temperature susceptibility is between 1.5 – 2.2 ($\times 10^{-5}$) emu/g and this range covers samples with 0 $<$ x $<$ 0.4 in the notation UO$_{2+x}$. A more recent study of the susceptibility in bulk stoichiometric (x = 0) single crystals \cite{gofryk_magnetization_2014} give values at 5 K between 1.4 and 1.8 ($\times 10^{-5}$) emu/g with a difference of about 5\% between the values for the field applied parallel to $<100>$ and $<111>$. This is in excellent agreement with the earlier work \cite{arrott_magnetic_1957}. Bulk UO$_2$ is known to become antiferromagnetic (AF) at 30.8 K with noncollinear ordering into a complex magnetic arrangement, together with ordering of the quadrupoles at the uranium site \cite{santini_multipolar_2009}, \cite{hurley_thermal_2022}, but the global symmetry of this AF configuration remains cubic. The AF moment magnitude at 5 K is 1.74 $\mu_{\mathrm{B}}$ \cite{faber_neutron_1976}. Furthermore, the electronic ground state has been known since the early (1989) measurements of the crystal-field in UO$_2$ \cite{amoretti_5_1989}, and confirmed in more recent experiments using synchrotron X-rays \cite{caciuffo_synchrotron_2023} and accompanying theory. These observations leave no doubt that the ground state is the $\Gamma_5$ triplet of the $^3H_4$ ground state, which can support a maximum magnetic ordered moment of 2 $\mu_{\mathrm{B}}$. 

Another important series of measurements are those reported by Jaime {\it et al}, \cite{jaime_piezomagnetism_2017} who applied magnetic fields of up to 60 {\it T} to bulk UO$_2$ over a range of temperatures. At no temperature (or magnetic field) was a large magnetic moment induced. Figure 3b of this paper shows the result of field sweeps up to $\pm$30 {\it T}, that resembles a straight line with a susceptibility of 7.12 m$\mu_{\mathrm{B}}/T$. This value corresponds to 1.471$\times$ 10$^{–5}$ emu/g, which is identical to the susceptibility measured by Arrott \& Goldman from polycrystalline samples in 1957 \cite{arrott_magnetic_1957}. There are, as discussed in \cite{jaime_piezomagnetism_2017}, important modifications in the AF structure of UO$_2$ with large fields applied along $<111>$, but the overall susceptibility of UO$_2$ is a robust parameter, as is the barrier against destroying the AF structure at low temperature. 

These properties listed above are, of course, relative to bulk UO$_2$. The question raised by Ref. \cite{sharma_induced_2022} is how much can the properties be changed by making thin films on perovskite substrates, which coherently strain the crystals and break the cubic symmetry? We argue that the strains of $\sim$ 1\% are insufficient to create a radical change in the properties. Our XMCD measurements support that opinion. Magnetic moments of over 3 $\mu_{\mathrm{B}}$ would seem highly unlikely given the electronic ground state known of bulk UO$_2$. Such moments are proposed in Fig. 5c of \cite{sharma_induced_2022} for UO$_2$/STO. In contrast, we do find a significant increase in the susceptibility of the UO$_2$ films (a factor of $\sim$ 2.5 at 5 K), but this is totally within the confines of our model for the magnetic behaviour of bulk UO$_2$. The fact that this increase in the susceptibility appears independent of the substrate, suggests that it may be more dependent on the thickness of the UO$_2$, and provides motivation for examining even thinner layers. Given the sensitivity of XMCD, this should be possible maybe even down to a few nanometers.

The measurement of the magnetization as a function of temperature (Fig. \ref{fig:3}) shows that the low-temperature susceptibility is a factor of about 2.5$\times$ greater than found in bulk AF UO$_2$. This suggests that the antiferromagnetic correlations in UO$_2$, which are present until at least 100 K \cite{caciuffo_magnetic_1999}, \cite{paolasini_anisotropy_2021}, are reduced in the thin film. Moreover, this is supported by the absence of any anomaly in the susceptibility in the temperature range around the ($T_{\mathrm{N}}$ = 30.8 K)  of bulk UO$_2$. A similar situation exists in UO$_{2.10}$ as shown in Fig$.$ \ref{fig:3}, taken from \cite{arrott_magnetic_1957}. In this latter material no ordering occurs, and the susceptibility at low temperature is also higher than in bulk UO$_2$. The absence of AF order in such a 20 nm film of UO$_2$ is consistent with the observations reported in \cite{bao_antiferromagnetism_2013} where a film of 24 nm of UO$_2$ on LAO was found to order, but no ordering was found in thinner films of 15 and 8 nm.

As mentioned in Sec. II, our films were capped with Nb, which was not the case in Ref. \cite{sharma_induced_2022}. The absence of a cap means that additional oxygen may be deposited at the surface when they are exposed to air. The (100) surface is known to be polar \cite{bottin_thermodynamic_2016} and the favoured re-arrangement to achieve charge neutrality with such a termination plane is with extra oxygens at the surface, although such perturbations only extend, according to theory, some $\sim$ 3 nm below the surface. These changes cannot be observed with reflectivity, as they involve only additional oxygen atoms, which scatter X-rays poorly, especially compared to uranium. 

Reference \cite{sharma_induced_2022} also reports polarized neutron reflectivity (PNR) experiments that appear to confirm the large moments in an 11 nm film of UO$_2$ on a YAO substrate (Fig. 6 of Ref. \cite{sharma_induced_2022}) at 10 K and an applied magnetic field of 4.8 {\it T}. If UO$_2$/YAO films have the same susceptibility as those in Table \ref{tab:table2} above, we should expect an induced moment of  $\sim$ 85 m$\mu_{\mathrm{B}}$. The manuscript reports a magnetisation of 11.64 emu/cm$^3$ and translates this to 210 m$\mu_{\mathrm{B}}$ per uranium atom. This is incorrect. The value of 210 m$\mu_{\mathrm{B}}$ per unit cell is correct, however, the unit cell of UO$_2$ contains 4 uranium atoms. Therefore, the value of the magnetisation per uranium atom would be equal to 51 m$\mu_{\mathrm{B}}$ per uranium. Given the uncertainties in the PNR determination, such a number is close to our estimate of 85 m$\mu_{\mathrm{B}}$ per U atom. Our modelling shows that the PNR results are consistent with a volume magnetization of 11.64 emu/cc, which suggests a field induced canting of the U moments leading to a small net magnetization.

\section{Conclusions}
The most important conclusion of these experiments is that we do not confirm the results of Ref. \cite{sharma_induced_2022}. Our results for the thin films can be explained within our present theoretical understanding of UO$_2$ whilst the results published in \cite{sharma_induced_2022} lie outside of such a theoretical understanding of this material, specifically in suggesting the antiferromagnetic coupling is weak, and in stating that magnetic moments well above the theoretical limit of 2 $\mu_B$/U atom can be induced by a relatively small magnetic field.

Although the investigated films remain paramagnetic (the XMCD technique is not directly sensitive to antiferromagnetism), they have a value of the magnetization somewhat larger than found in bulk UO$_2$. The M/H curves, however, are linear, quite different from reported in Ref. \cite{sharma_induced_2022}. Given the large difference between the substrate and UO$_2$ film thicknesses (25,000 if substrate is 0.5 mm), small effects in the substrates, such as reported by Khalid {\it et al.} \cite{khalid_ubiquity_2010} and Ney {\it et al.} \cite{ney_limitations_2008}, can easily be ascribed to large effects in the films. In studies such as this, it is important to use an element sensitive technique, such as XMCD.

The larger susceptibility (at low temperature) (Fig$.$ \ref{fig:3}) for a 20 nm film of UO$_2$ as compared to the bulk suggests that the antiferromagnetic correlations are reduced in the thin film. In turn, this is consistent with the observation that the magnetization has no anomaly at low temperature and such films do not order magnetically, at least above 5 K. 

\begin{acknowledgments}
This work was conducted under the Engineering and Physical Sciences Research Council (EPSRC), UK, Centre for Doctoral Training in Nuclear Energy Futures grant (EP/S023844/1), through the National Nuclear User Facility for Radioactive Materials Surfaces (NNUF-FaRMS), grant no. \textsc{EP/V035495/1}.
\end{acknowledgments}

\section{Appendix}
\begin{figure}
  \centering
 \includegraphics[width=1.0\columnwidth]{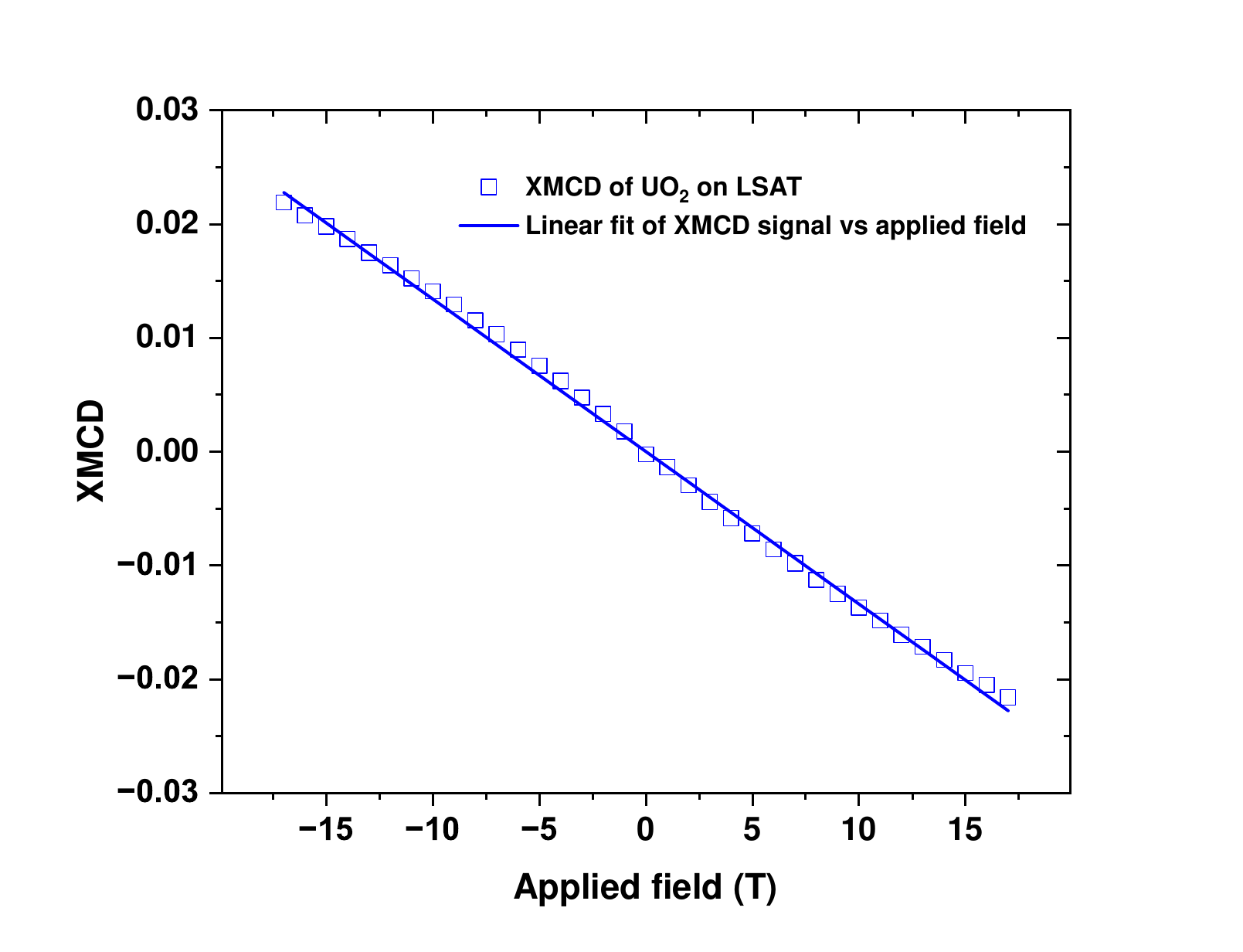}
 \caption{Amplitude of the XMCD signal at the uranium $M_4$ edge measured at 5 K as a function of the applied magnetic field B for a 20.3 nm thick UO$_2$ thin film deposited by DC magnetron sputtering on a La,Sr)(Al,Ta)O$_3$ (LSAT) substrate of 0.5 mm thickness and (001) orientation. The solid line is a linear fit to the data.}
  \label{A1}
  \centering
  \includegraphics[width=1.0\columnwidth]{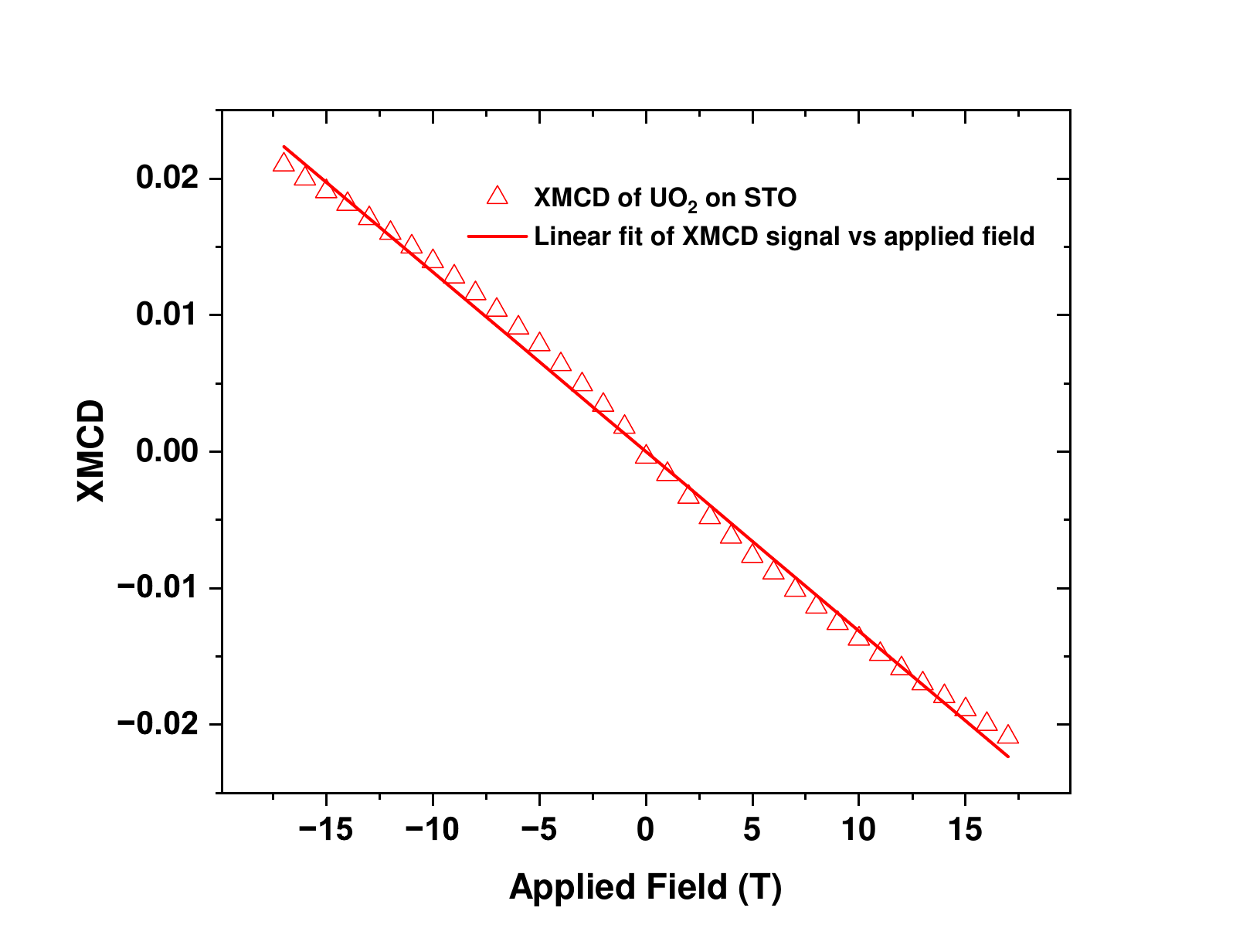}
  \caption{Amplitude of the XMCD signal at the uranium $M_4$ edge measured at 5 K as a function of the applied magnetic field B for a 20.6 nm thick UO$_2$ thin film deposited by DC magnetron sputtering on a SrTiO$_3$ (STO) substrate of 0.5 mm thickness and (001) orientation. The solid line is a linear fit to the data.}
  \label{A2}
\end{figure}

As the XMCD signal integrated over the $M_5$ edge is small, the amplitude of the XMCD signal at the uranium $M_4$ edge is a good representation of the magnetic moment induced on the uranium atoms by the applied magnetic field. Figs. \ref{A1} and \ref{A2} demonstrate the absence of any induced ferromagnetic order in the UO$_2$ thin films grown at the University of Bristol on (La,Sr)(Al,Ta)O$_3$ (LSAT) and SrTiO$_3$ (STO) substrates. 

\clearpage

\bibliography{UO2XMCD}
\end{document}